\documentclass[sigconf,natbib=false, nonacm]{acmart}

\AtBeginDocument{%
  }

\setcopyright{none}
\RequirePackage[
  datamodel=acmdatamodel,
  style=acmnumeric,
  ]{biblatex}

\begin{document}

\title{Beyond C/C++: Probabilistic and LLM Methods for Next-Generation Software Reverse Engineering}

\author{Zhuo Zhang}
\affiliation{%
  \institution{Purdue University}
  \city{West Lafayette}
  \state{IN}
  \country{USA}
}
\email{zhan3299@purdue.edu}

\author{Xiangyu Zhang}
\affiliation{%
  \institution{Purdue University}
  \city{West Lafayette}
  \state{IN}
  \country{USA}
}
\email{xyzhang@cs.purdue.edu}


\begin{abstract}
This proposal discusses the growing challenges in reverse engineering modern software binaries, particularly those compiled from newer system programming languages such as Rust, Go, and Mojo. 
Traditional reverse engineering techniques, developed with a focus on C and C++, fall short when applied to these newer languages due to their reliance on outdated heuristics and failure to fully utilize the rich semantic information embedded in binary programs.  
These challenges are exacerbated by the limitations of current data-driven methods,  which are susceptible to generating inaccurate results, commonly referred to as hallucinations.
To overcome these limitations, we propose a novel approach that integrates probabilistic binary analysis with fine-tuned large language models (LLMs). 
Our method systematically models the uncertainties inherent in reverse engineering, enabling more accurate reasoning about incomplete or ambiguous information. 
By incorporating LLMs, we extend the analysis beyond traditional heuristics, allowing for more creative and context-aware inferences, particularly for binaries from diverse programming languages. 
This hybrid approach not only enhances the robustness and accuracy of reverse engineering efforts but also offers a scalable solution adaptable to the rapidly evolving landscape of software development. 
\end{abstract}


\newcommand{\code}[1]{``\texttt{#1}''}



\maketitle

\section{Introduction}

In response to the growing demand for more flexible and secure programming practices~\cite{usrust}, recent years have seen the introduction of advanced system programming languages beyond traditional C and C++. 
Languages such as Go, Rust, Mojo, and even functional languages like OCaml have been developed, offering greater flexibility and safety for software development. 
However, the advantages and flexibility provided at the language level do not extend to software consumers once the software is distributed. 
Similar to C and C++, modern system programming languages compile code into native binaries, stripping away high-level abstractions and semantic information (e.g., Rust’s Traits), leaving only raw bytes. 
This loss of information presents significant challenges for software consumers who need to tailor the software to their specific needs, particularly for security purposes, such as patching vulnerabilities in commercial off-the-shelf (COTS) software~\cite{DBLP:conf/seke/LiXTMW17, DBLP:conf/icse/XuCCLS17}, debloating and customizing third-party libraries to reduce attack surfaces~\cite{DBLP:journals/corr/abs-1902-06570, DBLP:conf/dimva/Redini0MSVK19, DBLP:conf/acsac/AgadakosJWKP19, DBLP:conf/asplos/ZhangR0M22}, or maintaining and hardening legacy systems~\cite{DBLP:conf/ccs/CarboneCLLPJ09, DBLP:conf/uss/CarliniBPWG15, DBLP:conf/popl/MartinHCAC10}. 

To address this challenge, reverse engineering has become an essential discipline in software security, focusing on recovering source-level semantic information from compiled binaries. 
Over the past twenty years, there have been significant advancements in techniques such as disassembly~\cite{DBLP:conf/sp/YeZSAZ23, DBLP:conf/ndss/PeiGWYJ21, DBLP:conf/icse/MillerKSZZL19, DBLP:conf/ndss/BaumanLH18, DBLP:conf/eurosys/AltinayNKRZDGNV20, DBLP:conf/sp/ZhangYTALZ21}, function boundary identification~\cite{DBLP:conf/acsac/Alves-FossS19}, type inference~\cite{DBLP:conf/sp/ZhangYYTLKAZ21, DBLP:conf/ndss/SlowinskaSB11, DBLP:conf/sp/Shoshitaishvili16, DBLP:conf/sigsoft/PeiGBCYWUYRJ21,DBLP:conf/ndss/LinZX10, DBLP:conf/ndss/LeeAB11}, and the recovery of high-level abstractions~\cite{DBLP:conf/ccs/SchwartzCDGHH18}, including code structures~\cite{DBLP:conf/uss/BasqueBGOMBDS024}, data structures~\cite{DBLP:conf/ccs/SchwartzCDGHH18, DBLP:conf/sp/ZhangYYTLKAZ21}, and variable names~\cite{xu2023lmpa, pal2024len, DBLP:conf/uss/ChenLSGNV22, xie2024resym}.
Among these, probabilistic binary analysis~\cite{DBLP:phd/us/Zhang23} has emerged as a particularly promising technique. 
By leveraging probabilistic inference, it systematically models the inherent uncertainty caused by the aforementioned information loss in reverse engineering. 
This approach yields practical and efficient analysis results while offering a novel probabilistic guarantee of soundness.

Despite these successes, reverse engineering for modern software remains a challenging task in practice, particularly when dealing with software compiled from various system programming languages. 
For example, even the most advanced commercial reverse engineering tools, like IDA Pro, struggle to produce readable decompilation results for a simple ``Hello, World'' program written in Rust~\cite{idarust}. 
Similarly, the state-of-the-art type inference technique, Osprey~\cite{DBLP:conf/sp/ZhangYYTLKAZ21}, faces similar challenges.

In the following sections, we will first explore the unique challenges and emerging opportunities in reverse engineering modern software, especially when analyzing binaries compiled from languages beyond C and C++. We will then introduce our proposed techniques to address these challenges.

\section{Limitations of Existing Techniques}

We outline the key limitations of existing approaches as follows:

\smallskip
\noindent
\textbf{(L1) Most existing reverse engineering techniques are designed specifically for binaries compiled from C and C++.}  
Over the past two decades, reverse engineering has matured significantly and has seen widespread practical application. However, due to the dominance of C and C++ in earlier software development, nearly all reverse engineering tools and techniques, including those developed recently~\cite{DBLP:conf/sp/ZhangYYTLKAZ21}, are designed with binaries compiled from these languages in mind. 
Since compilation is an inherently lossy process, these techniques often require heuristic ``guesses'' to recover lost information, making them fundamentally dependent on heuristics during analysis (i.e., used for deterministic decision-making in conventional analysis or as probabilistic hints with prior probabilities in probabilistic analysis).
These heuristics, which may be effective for C and C++, are often incorrect or incomplete when applied to binaries compiled from newer programming languages.

\begin{figure}[t]
\includegraphics[width=0.47\textwidth]{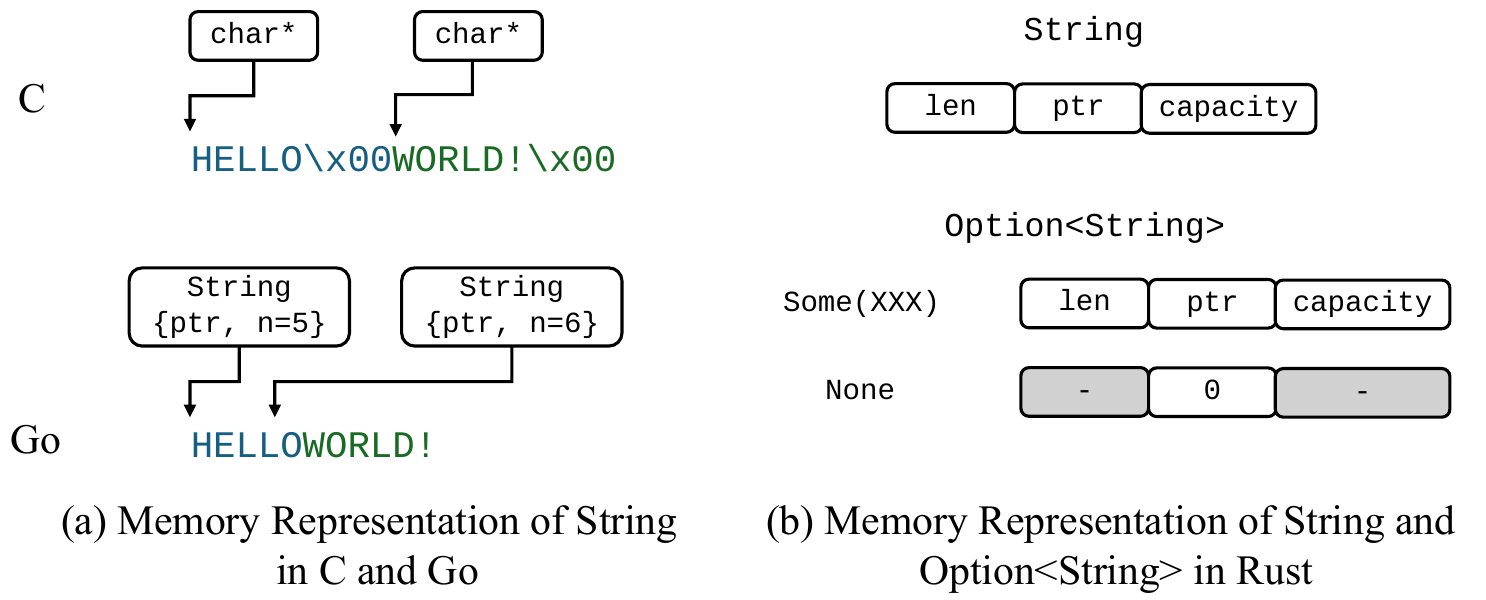}
\centering
\caption{Incorrect heuristics when applied to non-C languages.}
\label{fig:incorrectness}
\end{figure}

\smallskip \noindent \underline{Incorrectness.} Heuristics designed for C and C++ can produce entirely incorrect results when applied to binaries from other programming languages. 
Fig.~\ref{fig:incorrectness} demonstrates two such cases where these heuristics fail.
In Fig.~\ref{fig:incorrectness}(a), the memory layout of strings in both C and Go is compared. 
In C, strings are null-terminated, meaning that the end of a string is indicated by a null character (``\texttt{NULL}''). Consequently, each ``\texttt{char *}'' points to a sequence of characters ending with a null terminator. 
However, Go does not use null termination. 
Instead, Go represents a string using a structure with a ``\texttt{ptr}'' field, which points to the beginning of the string in memory, and an ``\texttt{n}'' field, which stores the string's length. 
As a result, Go programs do not contain null-terminated strings in memory.
The use of null characters to identify strings is a common heuristic in reverse engineering, aiding in tasks such as disassembly~\cite{DBLP:conf/icse/MillerKSZZL19}, variable name recovery~\cite{DBLP:conf/uss/ChenLSGNV22}, and binary-only fuzzing~\cite{DBLP:conf/sp/ZhangYTALZ21}. 
However, applying this heuristic to Go binaries, where strings are not null-terminated, will significantly reduce analysis accuracy.

An even more concerning issue is that some widely accepted principles in reverse engineering, including those trusted by human experts, may be fundamentally flawed. 
For example, in type inference, it is commonly assumed that if one composite structure is entirely and directly assigned the value of another, both structures should share the same type~\cite{DBLP:conf/ndss/LeeAB11, DBLP:conf/ndss/LinZX10, DBLP:conf/sp/ZhangYYTLKAZ21}. 
Note that type casting of entire composite structures is extremely rare.
In Fig.~\ref{fig:incorrectness}(b), we illustrate the memory layout of ``\texttt{String}'' and ``\texttt{Option<String>}'' in Rust. 
Specifically, a ``\texttt{String}'' is represented by three fields: ``\texttt{len}'', ``\texttt{ptr}'', and ``\texttt{capacity}'', which denote the length of the string, the memory address of the start of the string, and the number of bytes allocated for the string, respectively. ``\texttt{Option<String>}'', a wrapper type for ``\texttt{String}'', includes an additional value, ``\texttt{None}'', which indicates the absence of a string (otherwise wrapped by ``\texttt{Some}'', as in ``\texttt{Some(XXX)}'', denoting the wrapper of a string ``\texttt{XXX}'').
When examining the memory layout of ``\texttt{Option<String>}'', it is surprising to discover that ``\texttt{Some(XXX)}'' shares excatly the same memory layout as a simple ``\texttt{String}'', while ``\texttt{None}'' sets the ``\texttt{ptr}'' field to ``\texttt{0}''. 
It is important to note that Rust does not use null pointers, so programs can always differentiate between ``\texttt{Some}'' and ``\texttt{None}'' by checking whether the ``\texttt{ptr}'' field is ``\texttt{0}''.
Consider the statement ``\texttt{let b = Some(a)}'', where ``\texttt{a}'' is a ``\texttt{String}'' and ``\texttt{b}'' is hence an ``\texttt{Option<String>}''. 
At the binary level, there is a direct data flow from ``\texttt{a}'' to ``\texttt{b}'' without any modifications. 
Most existing type inference tools incorrectly assume that ``\texttt{a}'' and ``\texttt{b}'' are of the same type due to this strong heuristic. 
This becomes even more problematic, since ``\texttt{Option}'' is one of the most frequently used data structures in Rust, leading to widespread issues in type inference.

\begin{figure}[t]
\includegraphics[width=0.29\textwidth]{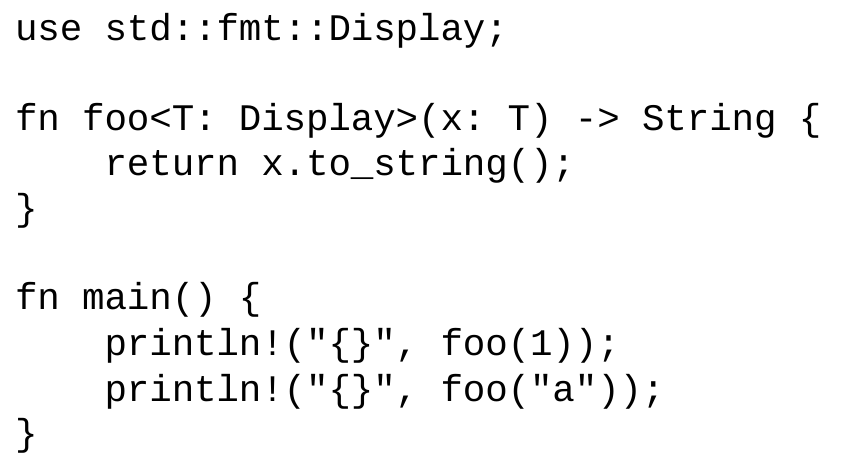}
\centering
\caption{Example of Traits in Rust}
\label{fig:incompleteness}
\end{figure}

\smallskip \noindent \underline{Incompleteness.} Additionally, newer languages introduce unique features, leading to the emergence of novel patterns in the compiled binaries. 
These patterns, which have the potential to significantly enhance reverse engineering efforts, are frequently overlooked by existing techniques.
A notable example is the concept of Traits in Rust. 
Traits in Rust are a way to define shared behavior across different types. 
They allow you to specify what functionalities a type must provide, enabling \textit{polymorphism} and code reuse.
One can think of Traits as Rust's innovative approach to implementing C++'s templates within its type system.
In Fig.~\ref{fig:incompleteness}, the ``\texttt{foo}'' function accepts a generic type ``\texttt{T}'' that must implement the ``\texttt{Display}'' trait, ensuring that the type can be converted into a string using the ``\texttt{to\_string()}'' method. 
The example demonstrates how the function ``\texttt{foo}'' can handle different types, such as an integer ``\texttt{1}'' and a string ``\texttt{``a''}'', both of which implement the Display trait. 
More importantly, when Rust code is compiled, Traits are \textit{monomorphized}. 
Specifically, in Fig.~\ref{fig:incompleteness}, the Rust compiler generates two distinct functions for ``\texttt{foo}'', one for integers and another for strings. 
From a reverse engineering perspective, given that these two functions are largely similar, aggregating information from both compiled functions could yield more reliable analysis results. 
Despite this, current techniques do not leverage this feature. 
One might argue that C++ templates exhibit similar behavior (which is also often overlooked by existing techniques). 
Unlike C++ templates, which are used sparingly due to their notorious complexity, Rust's Traits are seamlessly integrated into its type system and are widely employed in almost every Rust program. 
Leveraging Rust's Traits could significantly improve the accuracy of reverse engineering.
$\Box$

\smallskip \noindent \textbf{(L2) Most existing reverse engineering techniques overlook the \textit{descriptive semantics} of binary programs.} 
Programming languages encompass two levels of semantics: \textit{machine semantics}, which focus on how the program is executed by the machine, including data-flow and control-flow information, and \textit{descriptive semantics}, which reflect how the code is interpreted by human developers and analysts. 
Consider a binary search function where variable and function names are replaced with irrelevant ones (e.g., names associated with bubble sort). 
While this program might maintain its machine semantics, it loses its descriptive semantics, making it challenging for humans to comprehend.
It is because developers and analysts rely on both machine and descriptive semantics to understand and reason about programs. 
This reliance extends to reverse engineering, where analysts often depend not only on low-level data-flow and control-flow information but also on human-readable strings and recognizable constants (e.g., 2654435769, associated with the TEA encryption function~\cite{tea}), to form educated guesses~\cite{DBLP:conf/uss/MantovaniAFB22, DBLP:conf/uss/BurkPKV22}.
For instance, a function containing the string ``\texttt{Failed to connect to the target port}'' is likely network-related and may involve sockets. 
However, few existing techniques fully exploit descriptive semantics. 
Despite the potential of data-driven approaches to leverage descriptive semantics through machine learning's ability to understand natural language, these methods often mask strings and constants to reduce the token vocabulary size~\cite{DBLP:conf/sigsoft/PeiGBCYWUYRJ21, DBLP:conf/ndss/PeiGWYJ21}.
$\Box$

One could argue that with the recent advancements in Large Language Models (LLMs), which demonstrate superior capabilities in understanding natural language (and thus descriptive semantics) and in summarizing usable features encountered during training (and thereby deriving useful heuristics for emerging languages like Rust), LLMs could potentially be leveraged to address reverse engineering challenges.

\begin{figure}[t]
	\centering
	\includegraphics[width=0.98\linewidth]{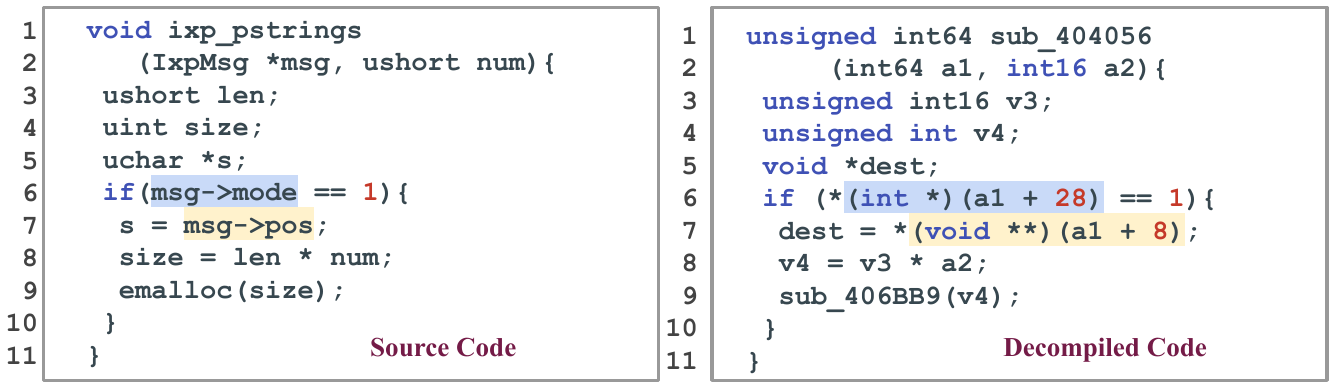}
	\caption{A 
 function's source code and decompiled code}
	\label{fig:motivating_decompiled}
\end{figure}

\begin{figure}[t]
	\centering
	\includegraphics[width=0.98\linewidth]{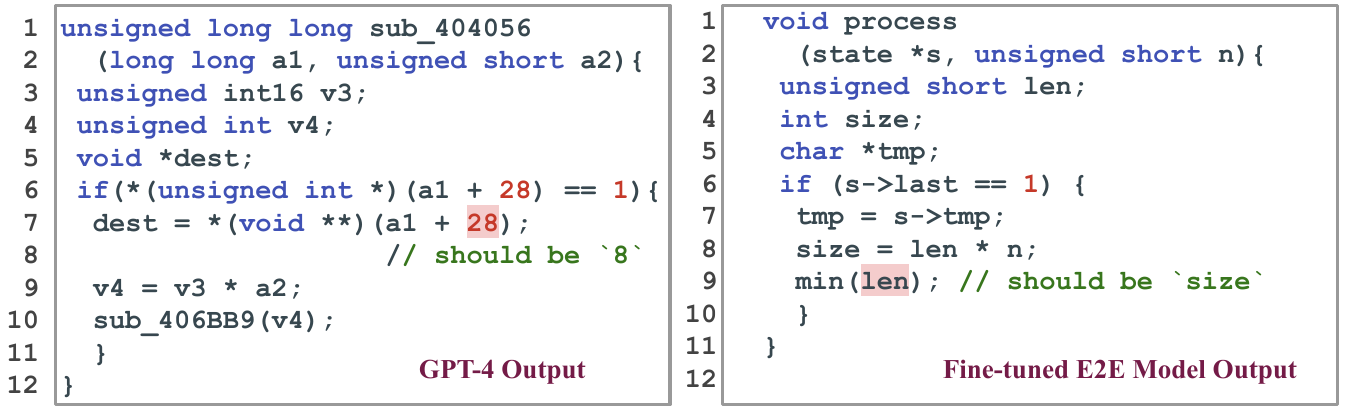}
	\caption{GPT-4 and direct fine-tuning are limited.} 
	\label{fig:motivating_gpt}
\end{figure}

\smallskip \noindent \textbf{(L3) Data-driven approaches are prone to hallucination and must be applied with caution in security-centric scenarios.} 
Recent studies~\cite{DBLP:conf/uss/0002TSAXLYW023} have shown that deep-learning-based approaches, including LLMs, are vulnerable to adversarial manipulation, where altering just a few bytes can entirely flip the analysis results. However, reverse engineering is often employed in security-sensitive contexts, such as binary hardening and malware analysis. Therefore, the results must be robust and, ideally, explainable to human analysts to ensure that final decisions are made with confidence.

To illustrate the limitations of LLMs, we present Fig.~\ref{fig:motivating_decompiled}, where we attempt to recover type and variable names from decompiled code generated by IDA Pro. 
The function accepts two parameters: a pointer \code{msg} to a user-defined structure \code{IxpMsg} and an unsigned short \code{num}. 
Lines 6 to 8 access two fields, \code{mode} and \code{pos}, of \code{msg}, and update two local variables, \code{size} and \code{s}. 

We initially explored using LLMs to recover this information directly. 
However, our efforts revealed the limitations of LLMs in this complex task. 
Simply prompting LLMs to enhance the readability of decompiled code led to suboptimal results. 
Fig.~\ref{fig:motivating_gpt} shows the output of GPT-4 when tasked with improving the readability of the decompiled code, i.e., in a zero-shot learning scenario. 
While it makes some adjustments (e.g., changing \code{(int *)(a1 + 28)} to \code{(unsigned int *)(a1 + 28)}), most variables remain unchanged, resulting in minimal improvement.
Moreover, GPT-4 fails to provide useful information about fields within user-defined structures, such as \code{IxpMsg}. 
This underscores a key challenge in using LLMs for reverse engineering: understanding decompiled code is inherently complex, typically requiring a human analyst with years of specialized training and adherence to well-established methodologies~\cite{DBLP:conf/uss/MantovaniAFB22, DBLP:conf/uss/BurkPKV22}. 
Expecting a general-purpose LLM to directly produce readable decompiled code is therefore unrealistic.

We also fine-tuned GPT-4 using an End-to-End (E2E) approach, where the input is decompiled code and the output is the corresponding source code. 
The goal was to enhance the model's proficiency in interpreting decompiled code. 
As shown in Fig.~\ref{fig:motivating_gpt}, this approach demonstrated some progress, such as accurately renaming \code{len}. 
However, it still falls short in providing comprehensive information and sometimes introduces errors, such as interpreting the field access \code{msg->mode} as \code{s->last}. 
Additionally, there is a significant issue with semantic divergence between the source code and the outputs of the fine-tuned model and GPT-4, including data-flow alterations (marked in red in Fig.~\ref{fig:motivating_gpt}). 
This divergence could potentially have substantial impacts on downstream security analysis.
$\Box$

This above discussion highlights the unique challenges and emerging opportunities in reverse engineering contemporary software, particularly when dealing with binaries compiled in languages beyond C and C++, in the context of the LLM era.

\section{Proposed Research}

\begin{figure}[t]
	\centering
	\includegraphics[width=0.98\linewidth]{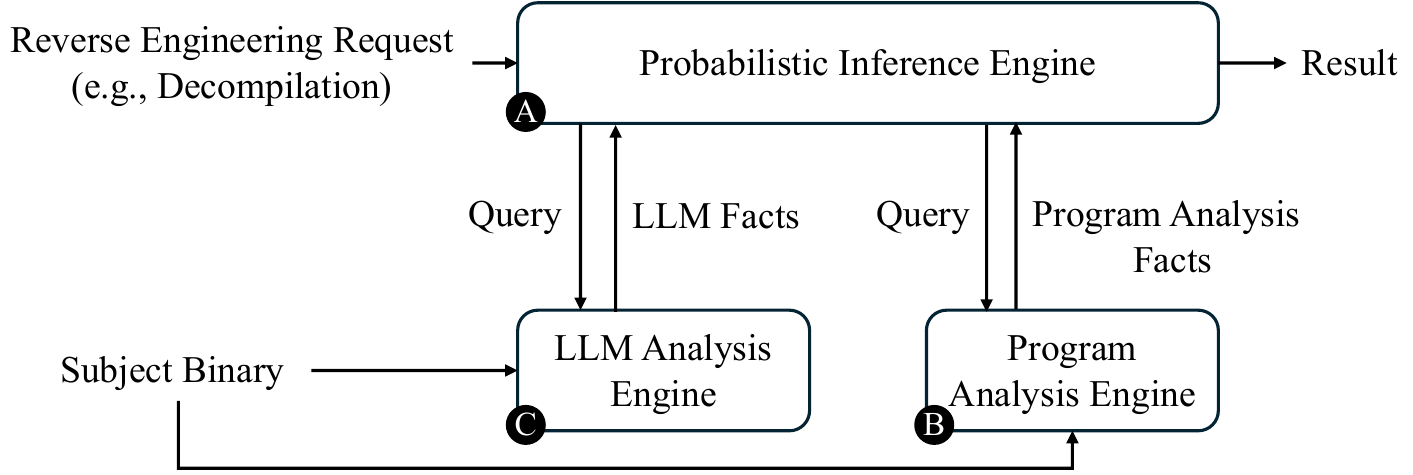}
	\caption{System design of the proposed technique.}
	\label{fig:arch}
\end{figure}

In this section, we introduce our proposed technique. Figure 5 illustrates the system design of our approach.
The overarching idea builds upon the success of probabilistic binary analysis~\cite{DBLP:phd/us/Zhang23}, where uncertainties in reverse engineering are systematically modeled, allowing for reasoning through probabilistic analysis. 
Specifically, the \textit{Program Analysis Engine} (Component B) examines the subject binary and collects facts, each accompanied by prior probabilities or confidence levels that indicate the likelihood of correctness. 
These analyzed facts are then fed into the \textit{Probabilistic Inference Engine} (Component A), where they are aggregated in a manner that accounts for their respective confidence levels.

However, as previously mentioned, existing probabilistic binary analysis techniques face two significant challenges: 1) limited heuristics that restrict analysis to C/C++ programs, and 2) an inability to leverage the descriptive semantics of subject binaries.
To address these challenges, we introduce the \textit{LLM Analysis Engine} (Component C). 
Rather than developing new heuristics for other languages, we leverage fine-tuned LLMs. 
It is important to note that creating high-quality heuristics for reverse engineering is a labor- and intelligence-intensive task, one that has taken researchers two decades to perfect for C and C++ programs. 
Given the rapid evolution of programming languages and compilers, we believe that utilizing LLMs offers a more effective solution. 
Moreover, LLMs can provide valuable insights into descriptive semantics.
In the sections, we will discuss each component in detail.

\begin{itemize}
    \item \textbf{Probabilistic Inference Engine.} 
    The Probabilistic Inference Engine is designed to systematically manage and reason about the uncertainties inherent in reverse engineering. 
    At a high level, it aggregates and evaluates facts derived from the analysis engines, assigning probabilities to these facts based on their confidence levels. 
    This approach allows the system to make informed decisions even in the presence of incomplete or ambiguous information. 
    Additionally, the Probabilistic Inference Engine plays a crucial role in mitigating potential biases introduced by heuristics from LLMs. 
    Heuristics, while powerful, can sometimes lead to overfitting or biased conclusions, particularly when the LLMs are applied to scenarios that differ from their training data (i.e., analyzing unseen malware). 
    By integrating these heuristics into a probabilistic framework, the engine can weigh them against other sources of information and adjust their influence accordingly. 
    This process helps to counterbalance any overconfidence or bias that might arise from the LLM-generated heuristics, ensuring that the final output remains robust and reliable, reflecting a balanced synthesis of all available data.
    
    \item \textbf{Program Analysis Engine.}
    Unlike existing probabilistic binary analysis techniques, which utilize program analysis for both heuristic guesses (e.g., identifying potential strings by detecting null terminators) and rigorous reasoning (e.g., data-flow analysis), our proposed approach takes a different direction. 
    In our technique, the Program Analysis Engine is dedicated solely to rigorous and methodical reasoning. 
    By removing the reliance on heuristic-based guesses, which can often introduce uncertainty, we ensure that the facts produced by this engine are grounded in well-founded analysis and have a higher degree of reliability.
    This exclusive focus on rigorous reasoning allows us to generate high-confidence data, which plays a critical role in the broader system. 
    High-confidence data from this engine serves as a robust counterbalance to the more heuristic and potentially error-prone outputs generated by the LLM Analysis Engine. This ensures that the final output of the system is not only comprehensive but also reliable and grounded in verifiable program analysis.

    \item \textbf{LLM Analysis Engine.}
    The LLM Analysis Engine, on the other hand, performs creative guesses, which can be applied to programs compiled from multiple programming languages. 
    This engine leverages the powerful pattern recognition and language understanding capabilities of LLMs to make educated guesses about code structures, semantics, and other aspects that may not be directly inferable through traditional program analysis. 
    By doing so, the LLM Analysis Engine extends the reach of our system to handle a diverse range of binaries, including those compiled from languages for which heuristic-based analysis might be limited or nonexistent.
    Furthermore, to fine-tune these LLMs, we propose using labels collected from source-code level analysis. 
    For example, labels can be derived by identifying relationships between functions in high-level languages, such as recognizing that two Rust functions have been monomorphized from the same generic function. 
    This process allows the LLMs to learn intricate patterns and nuances in code, enabling them to make more accurate inferences when faced with similar situations during analysis.
    As a result, LLMs can provide valuable insights that traditional analysis might overlook, effectively complementing the rigorous reasoning of the Program Analysis Engine. 
    This synergy offers a balanced approach that combines the creative flexibility of LLMs with the analytical precision of traditional program analysis.
\end{itemize}

\section{Conclusion}

In this proposal, we summarize the limitations of existing reverse engineering techniques, including dependency on outdated heuristics, challenges with descriptive semantics, and vulnerability to hallucinations in data-driven approaches.
To address this, we propose integrating probabilistic analysis with fine-tuned LLMs. This approach enhances accuracy and robustness, offering a more effective solution for analyzing binaries across diverse programming languages in security contexts.

\printbibliography

%

\end{document}